\documentclass[doublecol]{epl2} 
\usepackage[normalem]{ulem}   
\usepackage{epsfig}
\usepackage{amsmath}
\usepackage{amssymb}
\usepackage{mathrsfs}
\usepackage{chemarrow}
\usepackage{times}
\usepackage{graphicx}
\usepackage{xcolor}
\usepackage{epsfig,graphics}
\usepackage{color}
     \definecolor{darkred}{rgb}{0.75,0,0}
     \definecolor{darkgreen}{rgb}{0,0.5,0}
     \definecolor{darkblue}{rgb}{0,0,0.75}
     \definecolor{darkorange}{rgb}{1,0.9,0.1}

\newcommand{\rd}{\textrm{d}}
\def\mD{{\bf D}}
\def\vB{{\bf B}}
\def\vb{{\bf b}}

\def\vg {\mbox{\boldmath$\gamma$}}
\def\vx{{\bf x}}
\def\vu{{\bf u}}
\def\vv{{\bf v}}

\def\vX{{\bf X}}
\def\vJ{{\bf J}}

\title{Kinematic Basis of Emergent Energetics of Complex Dynamics}

\author{Hong Qian \and Yu-Chen Cheng  \and Ying-Jen Yang }
\shortauthor{H. Qian \etal}

\institute{   Department of Applied Mathematics, University of Washington,
Seattle, WA 98195-3925, USA
}

\abstract{
Stochastic kinematic description of a complex dynamics is
shown to dictate an energetic and thermodynamic structure.
An energy function $\varphi(\vx)$ emerges as the limit of the generalized, nonequilibrium free energy of a Markovian dynamics
with vanishing fluctuations.  In terms of the $\nabla\varphi$ 
and its orthogonal field $\vg(\vx)\perp\nabla\varphi$,
a general vector field $\vb(\vx)$ can be decomposed into $-\mD(\vx)\nabla\varphi+\vg$,
where $\nabla\cdot\big(\omega(\vx)\vg(\vx)\big)
=-\nabla\omega\mD(\vx)\nabla\varphi$.  The matrix
$\mD(\vx)$ and scalar $\omega(\vx)$, two additional 
characteristics to the $\vb(\vx)$ alone, represent the local 
geometry and density of states intrinsic to the statistical motion 
in the state space at $\vx$.  $\varphi(\vx)$ and $\omega(\vx)$ 
are interpreted as the emergent energy and 
degeneracy of the motion, with an energy balance equation
$\rd\varphi(\vx(t))/\rd t=\vg\mD^{-1}\vg-\vb\mD^{-1}\vb$,
reflecting the geometrical $\|\mD\nabla\varphi\|^2+\|\vg\|^2=\|\vb\|^2$.  The partition function employed in statistical mechanics and 
J. W. Gibbs' method of ensemble change naturally arise; 
a fluctuation-dissipation theorem is established via the two 
leading-order asymptotics of entropy production as $\epsilon\to 0$. 
The present theory provides a mathematical
basis for P. W. Anderson's emergent behavior in the hierarchical
structure of complexity science. }
\begin{document}
\maketitle

\section{Introduction}

	Classical mechanics has traditionally been divided into
{\em kinematics} and {\em dynamics}.  The former gives the
precise relationship between a mechanical motion $\vx(t)$
and descriptions of the motion in terms of its 
velocity $\dot{\vx}(t)$ and acceleration 
$\ddot{\vx}(t)$ under geometric constraints, and the latter 
provides relationships between the motions and the concepts 
of mass, force, and \textit{ mechanical energy}.  While the former
is a part of calculus, the latter constitutes the core of the 
classical physics of motion.  When a mechanical system contains 
a great many number of point masses such as atoms and 
molecules, the notions of {\em heat} and {\em temperature}, 
as a stochastic description of complex mechanical 
motions and kinetic energy, arise.  

	In classical, macroscopic chemical kinetics, a reaction in 
aqueous solution, say $A+B\rightarrow C$, is described by a rate process: $\rd c_A(t)/\rd t = -r(t)$, which should be identified as
the chemical kinematics.  It is again based on calculus which 
rigirously defines the concept of instantaneous rate of 
concentration change (fluxion). The functional relationship between
$r(t)$ and the concentrations $c_A(t)$ and $c_B(t)$, known as 
a {\em rate law}, is analogous to the constitutive relations.  One 
well-known example of a rate law is the Guldberg-Waage 
mass action $r(t)=kc_A(t)c_B(t)$
where $k$ is a constant independent of $c_A$ and $c_B$. 
The notion of energy in chemical reactions, however, exists 
in the chemical thermodynamics of heterogeneous substants, 
a separate theory developed 
by J. W. Gibbs, who introdued the notion of {\em Gibbs function} 
and {\em chemical potential} as the energy and force 
that drives the chemical changes.
 
	With the above understandings, therefore, it came as a surprise
that a recent work \cite{ge-qian-16-1,ge-qian-16-2} claims 
that the mathematical foundation of Gibbsian chemical 
thermodynamics, at a given temperature, 
needs only the mesoscopic stochastic 
kinematics, irrespective of any details of the mechanics of
the atoms and molecules.  In other words, the isothermal 
chemical thermodynamics of Gibbs is dictated purely by 
the kinematics via the mathematics of probability.  The 
implication of this observation is conceptually sigificant: It
implies in general a stochastic description of a complex dynamics 
has a ``hidden'' energetics that is already defined by 
mathematics!  The present paper applies this novel
idea, {\em stochastic kinematics dictates energetics}, to 
another general class of processes and further explore the 
idea: diffusion processes in continuous space $\mathbb{R}^n$ 
with continuous time $t\in\mathbb{R}$.  Mathematical analysis 
again reveals a hidden energetic and thermodynamic structure 
that underlying the kinematics.  The term ``thermo'' here does not imply heat; rather it means a stochastic 
description of a complex dynamics. 

The concept of {\em thermodynamic force} was clearly 
articulated in the work of L. Onsager \cite{onsager-31}. 
In chemical kinetics, it is widely accepted that {\em entropic
force} is a legitimate description on par with mechanical force 
as a ``cause'' for an action \cite{kirkwood-35,hill-book}.  
The Shannon entropy is computable 
in any statistical description of dynamics \cite{khinchin,mackey-rmp}. 
Actually, P. W. Anderson has stated that
``[A]t each level of complexity entirely new properties appear, 
and the understanding of the new behaviors requires research
which I think is as fundamental in its nature as any other.''
\cite{pwa}  In fact, he continued to provide a recipe for
discovering an emergent law:
\begin{quote}
``It is only as the nucleus is considered
to be a many-body system | in what is
often called the $N\to\infty$ limit | that such
 [emergent] behavior is rigorously definable.  ... 
Starting with the fundamental
laws and a computer, we
would have to do two impossible things
| solve a problem with infinitely many
bodies [e.g., zero fluctuations], and then apply the result to a
finite system | before we synthesized
this behavior.''
\end{quote}
As we shall see, the discovery of the hidden thermodynamic
laws indeed involves taking the limit of noise, e.g., 
fluctuations, tending zero.
Our result thus provides a clear understanding of why and how
emergent thermodynamic behaviors, as statistical laws, can
be independent of the underlying details.

	We consider the general description of complex
dynamics in terms of a stochastic Markov process $\vX_{\epsilon}(t)$.
Such a dynamics can be represented by its probability distribution $p_{\epsilon}(\vx,t)$ that follows a Fokker-Planck equation (FPE)
\cite{gardiner}
\begin{equation}
  \frac{\partial p_{\epsilon}}{\partial t} 
     = -\nabla\cdot\vJ\big[p_{\epsilon}\big], \
   \vJ\big[p_{\epsilon}\big] \equiv \vb(\vx)p_{\epsilon}
             -\epsilon \mD(\vx) \nabla p_{\epsilon}.
\label{eq1}
\end{equation}
Its trajectory is represented in terms of the solution to a Langevin type equation 
$\rd\vX_{\epsilon}(t) = \tilde{\vb}(\vX_{\epsilon})\rd t + \big[2\epsilon \mD(\vX_{\epsilon})\big]^{\frac{1}{2}}\rd \vB(t)$, where $\vB(t)$ is the standard multidimentional Brownian motion and $\tilde{\vb}=\vb+\nabla \cdot \epsilon\mD$.
The $\epsilon$ signifies a connection 
between deterministic and stochastic motions \cite{qian-epjst}: In the limit of
$\epsilon\to 0$, the stochastic trajectory $\vX_{\epsilon}(t)\to
\hat{\vx}(t)$ which satisties $\rd\hat{\vx}(t)/\rd t = \vb\big(
\hat{\vx}(t)\big)$, and similarly $p_{\epsilon}(\vx,t)\to
\delta\big(\vx-\hat{\vx}(t)\big)$ if initial value $p_{\epsilon}(\vx,0)=\delta\big(\vx-\hat{\vx}(0)\big)$.

	We emphasize that a stochastic description of dynamics does not
have any notion of ``energy'' and ``thermodynamics'' at the onset.
This is what we meant by ``kinematic description''.  As we shall 
show, however, in addition to the actual limit $\hat{\vx}(t)$, 
a mathematico-thermodynamic structure also emerges in the 
process of taking $\epsilon\to 0$:  For a fixed $\epsilon$ and 
regarding the stochastic dynamics in Eq. \eqref{eq1}, it is known that the relative entropy 
\begin{equation}
\label{the-F}
    F\big[p_{\epsilon}(\vx,t)\big] \equiv \int_{\mathbb{R}^n}
               p_{\epsilon}(\vx,t)\ln\left(\frac{p_{\epsilon}(\vx,t)}{\pi_{\epsilon}(\vx)}
          \right)\rd\vx,
\end{equation}
has a paramount importance \cite{lebowitz-bergmann,qian-01-free-energy,info-book}, where the stationary solution to \eqref{eq1}
$\pi_{\epsilon}(\vx)$ embodies the notion of an entropic force.  It 
was discovered only recently that the $F$ satisfies a 
balance equation $\rd F/\rd t \equiv -f_{\mathrm{d}}(t) = Q_{\mathrm{hk}}(t)-e_{\mathrm{p}}(t)$
\cite{ge-qian-10,vandenbroek-esposito,esposito-vandenbroek,qian-jmp}, with
\begin{subequations}
\label{3q}
\begin{eqnarray}
   f_{\mathrm{d}}\big[p_{\epsilon}\big] &=&  \int_{\mathbb{R}^n}
          \vJ\big[p_{\epsilon}\big] \cdot \nabla
            \ln\left(\frac{p_{\epsilon} }{\pi_{\epsilon} } \right) \rd\vx,
\\
	Q_{\mathrm{hk}}\big[p_{\epsilon}\big] &=&  \int_{\mathbb{R}^n}
          \vJ\big[p_{\epsilon}\big]\cdot(\epsilon\mD)^{-1}(\vx)\vJ\big[\pi_{\epsilon}\big]
            \pi^{-1}_{\epsilon}\rd\vx, \quad
\\
	e_{\mathrm{p}}\big[p_{\epsilon}\big] &=& \int_{\mathbb{R}^n}
               \vJ\big[p_{\epsilon}\big]\cdot(\epsilon\mD)^{-1}(\vx)\vJ\big[p_{\epsilon}\big]
            p^{-1}_{\epsilon}\rd\vx.
\end{eqnarray}
\end{subequations}
All three quantities in Eq. \eqref{3q} are non-negative. This fact ensures 
the interpretation of $Q_{\mathrm{hk}}$ and $e_{\mathrm{p}}$ as the source and 
sink for the ``free energy'' of a system. The $F\big[p_{\epsilon}\big]$ 
is interpreted as a generalized, nonequilibrium free energy since 
$-\epsilon\int_{\mathbb{R}^n} p_{\epsilon}\ln\pi_{\epsilon}\rd\vx$,
$-\int_{\mathbb{R}^n} p_{\epsilon}\ln p_{\epsilon}\rd\vx$, and
$\epsilon$ are analogous to ``mean internal energy'' $E$, 
entropy $S$, and temperature $T$, thus $F=E-TS$.
See \cite{jarzynski,seifert,esposito-review} for the relation between these average quantities and the trajectory-based stochastic thermodynamics,
Jarzynski like equalities, and fluctuation theorems.

The key result of the present work is the physical interpretation
of a set of three equations in Eqs. \eqref{hq-eq} putting together
as a system. The three equations were known but not with the thermodynamic connections and the geometric interpretation shown in this letter.  Eqs. \ref{hq-eq}a,b were in the mathematical 
work of Ventsel and Freidlin and the studies that followed
\cite{freidlin-wentzell,dykman,graham-tel}.
(\ref{hq-eq}c) has been derived in the work of Graham and T\'{e}l \cite{graham-tel-JSP}. They focused on constructing asymptotic stationary probability distributions 
based on the three equations.  

The present work illustrates a deep relation between this set of
equations as a dynamic description and thermodynamic energetics, 
in the context of the recently developed stochastic thermodynamics. 
We regard Eq. \eqref{hq-eq} as an emergent {\em energetics} that accompanies the {\em deterministic dynamics} $\dot{\vx}=\vb(\vx)$.  As a characterization 
of a complex behavior, the $\vb(\vx)$ does not exist alone as a 
macroscopic description, it comes with additional
information coded in $\mD$, $\varphi$, and $\omega$, as {\em thermodynamics}.  The Pythagorean relation 
\eqref{p-like-eq} among the key quantities in \eqref{hq-eq} further suggests a geometric perspective to be explored.

We note the Eq. (\ref{hq-eq}a) is not 
a Helmoholtz (or Hodge) decomposition of a vector field
since $\vg(\vx)$ is not divergence free  \cite{qian-pla}.
The motion following $\vg(\vx)$ however, conserves the
$\varphi(\vx)$ according to Eq. (\ref{hq-eq}b).
The divergence of $\vg(\vx)$ is provided by the
third equation (\ref{hq-eq}c).  The emergence of 
the $\omega(\vx)$ indicates the significance of a ``local volume'' 
of this $\varphi$-conservative motion
\cite{ge-qian-chaos}: This is called degeneracy in the
classical statistical mechanical termonology.  While
the motion following $\vg(\vx)$ can be complex, a proper
statistical description, called physical measure,  
usually exists \cite{lsyoung,ma-ao}.
For more discussion of the thermodynamic meanings of 
dissipative vs. conservative motions, and their relation to 
trajectory-based entropy production, see \cite{qian-pla,qian-epjst}.

\section{Emergent potential and $\varphi$-conservative
motion $\vg$}

The mathematical theory of large deviations connects 
the stochastic dynamics with finite $\epsilon$ to the 
deterministic dynamics with $\epsilon=0$ \cite{freidlin-wentzell-book,dembo-book,oono-LDT,touchette,eric-smith-LDT,qian-ijmpb}.  It is a rigorous and complete asymptotic theory akin to the WKB ansatz:
\begin{equation}
\label{wkb}
    p_{\epsilon}(\vx,t) = \exp\big[-\varphi^{\mathrm{td}}(\vx,t)/\epsilon
         +  o\big(\epsilon^{-1}\big) \big],
\end{equation}
in which $\varphi^{\mathrm{td}}(\vx,t)$ is known as a time-dependent
{\em large deviation rate function}. \footnote{The $\varphi^{\mathrm{td}}(\vx,t)$ satisfies a time-dependent equation of its own: $\partial\varphi^{\mathrm{td}}(\vx,t)/\partial t =-\nabla\varphi^{\mathrm{td}}(\vx,t)\cdot\vg^{\mathrm{td}}(\vx,t)$,
where $\vg^{\mathrm{td}}(\vx,t)\equiv\mD(\vx)\nabla\varphi^{\mathrm{td}}(\vx,t)+\vb(\vx)$.  This is known as a Hamilton-Jacobi equation.}  Taking the Eq. \eqref{wkb} as given
and recall that $p_{\epsilon}(\vx,t)\to
\delta\big(\vx-\hat{\vx}(t)\big)$, then it is easy to see that 
in the limit of $\epsilon\to 0$,
\begin{equation}
\label{eq5}
   \epsilon F\big[p_{\epsilon}(\vx,t)\big] 
  \to \varphi^{\mathrm{ss}}\big(\hat{\vx}(t)\big), \
  \varphi^{\mathrm{ss}}(\vx) = -\lim_{\epsilon\to 0} \epsilon\ln
        \pi_{\epsilon}(\vx).
\end{equation}
Since $\rd F/\rd t = -f_{\mathrm{d}}\le 0$,
$F\big[p_{\epsilon}(\vx,t)\big]$ is a monotonic
non-incresing function of $t$.  Consequently, Eq. \eqref{eq5}
states that $\varphi^{\mathrm{ss}}\big(\hat{\vx}(t)\big)$ is
a monotonic function of $t$, which implies that 
$\varphi^{\mathrm{ss}}(\vx)$ is an energy function of the dissipative
dynamics $\hat{\vx}(t)$, the solution to the 
deterministic equation $\rd\vx/\rd t = \vb(\vx)$.
We shall drop the superscript from the steady-state
large deviation rate function $\varphi^{\mathrm{ss}}(\vx)$ 
from now on. 

	The connection between $F[p_{\epsilon}]$ and 
$\varphi(\vx)$ illustrates that the latter is an emergent
energetic quantity in the limit of $\epsilon\to 0$.
It provides a new derivation with thermodynamic insights
of the mathematical result of Ventsel and Freidlin \cite{freidlin-wentzell,dykman,graham-tel}.  The relation
firmly connects (mesoscopic)
stochastic free energy $F[p_{\epsilon}]$ with 
(macroscopsic) deterministic pseudo-potential
$\varphi(\vx)$; it validates the earlier interpretation
of $-\epsilon\int_{\mathbb{R}^n} p_{\epsilon}\ln\pi_{\epsilon}\rd x$
as a mean internal energy, with $\varphi(\vx)$ as
the internal energy function of state $\vx\in\mathbb{R}^n$.

We emphasize that the
original $\vb(\vx) $ in Eq. \eqref{eq1} is not a gradient field in
general.  In terms of the newfound $\varphi(\vx)$ and 
continue the WKB ansatz, one can express 
$   \pi_{\epsilon}(\vx)=\omega(\vx)\exp[
    -\varphi(\vx)/\epsilon 
      +a\ln\epsilon + O(\epsilon)]$, 
in which $\ln\omega(\vx)$ is the next order to the leading
two terms and $a\ln\epsilon$ arising from normalization fractor
is independent of $\vx$.   Substituting this expression into  the stationary FPE 
\begin{equation}
 -\nabla \cdot \vJ[\pi_{\epsilon}(\vx)]=\nabla\cdot \big( \epsilon\mD(\vx)\nabla\pi_{\epsilon}(\vx)
          - \vb(\vx)\pi_{\epsilon}(\vx) \big) = 0,
\label{eq6}
\end{equation}
one obtains
$$
   \epsilon^{-1}\omega(\vx)\vg(\vx)\cdot\nabla\varphi -  
   \big[ \nabla\omega(\vx)\cdot\mD(\vx)\nabla\varphi + 
            \nabla\cdot\big(\omega(\vx)\vg(\vx)\big) \big]
$$
\begin{equation}
   +\epsilon\nabla\cdot\Big(\mD(\vx)\nabla\omega(\vx)\Big) 
     + O(\epsilon) = 0,
\label{eqn7}
\end{equation}
in which $\vg(\vx)\equiv \mD(\vx)\nabla\varphi(\vx)+\vb(\vx)$.
Equating like order terms in Eq. \eqref{eqn7}, we have
\begin{equation}
    \nabla\varphi(\vx)\cdot\vg(\vx) = 0,
       \  \forall \vx\in\mathbb{R}^n,
\label{orthog-eq}
\end{equation} 
the vector field $\vg(\vx)$ is orthogonal to
$\nabla\varphi(\vx)$, and \cite{graham-tel-JSP}
\begin{equation}
 \nabla\cdot\big( \omega(\vx)\vg(\vx) \big) 
    = -\nabla\omega(\vx) \cdot \mD(\vx)\nabla\varphi(\vx).
\label{eq-15}
\end{equation}
Actually, $\vg_{\epsilon}(\vx)\equiv$
$\pi^{-1}_{\epsilon}(\vx)\vJ[\pi_{\epsilon}(\vx)]$
has been identified as Onsager's thermodynamic 
flux\cite{qian-2002-iii}, and 
$\gamma$ is its limit as $\epsilon\to 0$:
\begin{eqnarray}
    \lim_{\epsilon\to 0}
     \vg_{\epsilon}(\vx) =
 \vb(\vx) + \mD(\vx)\nabla\varphi(\vx) \equiv \vg(\vx).
\label{decomp-eq}
\end{eqnarray}
The motions following the vector field $\vg(\vx)$ is restricted
on the level set of $\varphi(\vx)$.  This result also has a 
correspondence when $\epsilon$ is finite \cite{haken-graham,qqt}:
The stationary FPE can be re-written as
\begin{equation}
  \nabla\varphi_{\epsilon}(\vx)\cdot
            \vg_{\epsilon}(\vx) =
              \epsilon\nabla\cdot\vg_{\epsilon}(\vx),
\label{eq-34}
\end{equation} 
in which $\varphi_{\epsilon}(\vx)\equiv-\epsilon\ln\pi_{\epsilon}(\vx)$ has been widely called a kinetic potential \cite{haken-graham,kubo,nicolis-lefever}.  Therefore, if $\nabla\cdot\vg_{\epsilon}=0$ $\forall\epsilon$, then 
there is an orthogonality between $\nabla\varphi_{\epsilon}(\vx)$
and $\vg_{\epsilon}(\vx)$ for all $\epsilon$. 
More generally, irrespective of $\nabla\cdot\vg_{\epsilon}$ being zero or not, in the limit of $\epsilon\to 0$, $\varphi_{\epsilon}(\vx)\to\varphi(\vx)$,
$\vg_{\epsilon}\to\vg$, and $\vg\cdot\nabla\varphi=0$.

	$\vg(\vx)=0$ is mathematically equivalent to detailed balance. 
Stochastic systems with $\vg=0$ is widely considered
as ``non-driven'' \cite{zqq,gqq} and is expected to approach to an 
equilibrium steady state in the long-time limit.  For such systems,
the free energy $F$ aquires additional meaning as the
potential of Onsager's thermodynamics force,
$\vg_{\epsilon}=\mD(\vx)\nabla F$.

Collecting Eqs.  \eqref{orthog-eq}, \eqref{eq-15}
and \eqref{decomp-eq}, we have a system of three
equations
\begin{subequations}
\label{hq-eq}
\begin{eqnarray}
  & \vb(\vx) =  -\mD(\vx)\nabla\varphi(\vx) + \vg(\vx),
\\[3pt]
   &  \nabla\varphi(\vx)\cdot\vg(\vx) = 0,
\\[3pt]
  &\nabla\cdot\big( \omega(\vx)\vg(\vx) \big) 
    = -\nabla\omega(\vx) \cdot \mD(\vx)\nabla\varphi(\vx),
\end{eqnarray}
\end{subequations}
in which the vector fields $\vb(\vx)$ and $\vg(\vx)$ represent dynamics,
$\mD(\vx)$, which represents stochastic motion, can be
thought as a geometric metrics, 
$\omega(\vx)$ represents local ``measure'' for
the state space volume (degeneracy in the
statistical mechanical termonology)\footnote{Actually, 
there is also a time-dependent equation for $\omega^{\mathrm{td}}(\vx,t)$:
$\partial\omega^{\mathrm{td}}(\vx,t)/\partial t=
-\big(\mD(\vx)\nabla\varphi^{\mathrm{td}}(\vx,t)\big)\cdot\nabla\omega^{\mathrm{td}}(\vx,t)-\nabla\cdot\big(\vg^{\mathrm{td}}(\vx,t)\omega^{\mathrm{td}}(\vx,t)\big)$.  
The asymptotic expansion to this
next-order also provides a natural viscosity solution for the HJE
with a diffusion term $\epsilon\nabla\cdot\big(\omega^{\mathrm{td}}(\vx,0)\mD(\vx)\nabla\varphi^{\mathrm{td}}(\vx,t)\big)$.},
$\varphi(\vx)$ and $\ln\omega(\vx)$
are thermodynamic quantities akin to
energy and entropy, respectively. 
The ``noise structure'' $\mD(\vx)$
provides a unique geometry for the dynamics.  

In the simplest case, if $\mD(\vx)$ is the identity matrix, and 
$\omega(\vx) = 1$ is 
the Lebesgue measure, then the equations in Eq.
\eqref{hq-eq} become
\begin{subequations}
\label{h-sys}
\begin{eqnarray}
	&	\vb(\vx) = -\nabla\varphi(\vx) +\vg(\vx),
\\[3pt]
    &   \vg(\vx)\cdot\nabla\varphi(\vx) = 0, 
\\[3pt]  
    &  \nabla\cdot \vg(\vx) = 0. 
\end{eqnarray}
\end{subequations}
The vector field $\vg(\vx)$ is now volume preserving, and the system $\vx'(t)=\gamma(\vx)$ also has a conserved quantity $\varphi(\vx)$.
System with Eq. \eqref{h-sys} is intimately related to the
classical Hamiltonian systems \cite{perko}.
One of the most important features of this class of
dynamics is that the $\varphi(\vx)$ gives the 
steady state probability distribution exactly
for any finite $\epsilon$ in the form of
$\pi_{\epsilon}(\vx)\propto e^{-\varphi(\vx)/\epsilon}$
according to Boltzmann's law, if one identifies
$\epsilon$ with temperature \cite{qian-pla,qian-epjst}.

\section{$\varphi$-based statistical mechanics and ensemble
change}
\label{sec:3}

	It is seen immediately that if one computes a 
partition function from energy function $\varphi(\vx)$ and
degeneracy $\omega(\vx)$:
\begin{equation}
     Z(\epsilon) = \int_{\mathbb{R}^n} \omega(\vx)e^{-\varphi(\vx)/
              \epsilon}\rd\vx, 
\label{pf}
\end{equation}
then $Z^{-1}(\epsilon)\omega(\vx)e^{-\varphi(\vx)/\epsilon}$
is the asymptotic probability density for $\pi_{\epsilon}(\vx)$.
If the principle of equal probability is valid, e.g., 
$\nabla\cdot\vg(\vx)=0$, then it is the stationary
probability density of Eq. \eqref{eq1} for all $\epsilon>0$.

	Let us now consider a bivariate stochastic dynamics
with $\vx$ and $y$ which is assumed to be a scalar for simplicity.
The stationary joint probability for $\vx$ and $y$, $p_{\epsilon}(\vx,y)$ is related to the stationary conditional probability $p_{\epsilon}(\vx|y)$
through the marginal distribution for variable $y$,
$p_{\epsilon,y}(y)$: 
$p_{\epsilon}(\vx,y)=p_{\epsilon}(\vx|y)p_{\epsilon,y}(y)$.
In the asmptotic limit of $\epsilon\to 0$, this yields 
\begin{equation}
      \varphi(\vx,y) = \varphi(\vx|y) +  \varphi_{y}(y),
\end{equation}
and the partition functions
\begin{equation}
	Z_{\vx,y}(\epsilon) =   \int_y  Z_{\vx|y}(\epsilon;y)
         e^{-\varphi_y(y)/\epsilon} \rd y.
\label{ench-1}
\end{equation}
The $Z_{\vx|y}$ is the partition function with fixed
$y$, treated as a parameter, and $Z_{\vx,y}$ is the partition 
function with fluctuating $y$.  To asymptotically evaluating 
the integral in Eq. \eqref{ench-1}, it can be shown that at 
$y=\overline{y}$, the mean value of the fluctuating $y$:
\begin{equation}
       \frac{\partial}{\partial y}
 \varphi_y(\overline{y}) = \epsilon \left[ \frac{\partial}{\partial y} 
 \ln Z_{\vx|y}(\epsilon;y) \right]_{y=\overline{y}}
     \equiv \xi_y,
\label{ench-2}
\end{equation}
where $\xi_y$ is the conjugate variable to $y$.
Therefore, the integrand in Eq. \eqref{ench-1} can be approximately
expressed as
\begin{equation}
     \varphi_y(y) \simeq  \varphi_y(\overline{y}) + 
             \xi_y \big(y-\overline{y}\big).
\label{ench-3}
\end{equation}
Eqs. \eqref{ench-1}, \eqref{ench-2}, and \eqref{ench-3} constitute
 J. W. Gibbs' theory of ensemble change.  In doing so, the
thermodynamics of a stationary system with fluctuating $y$ and the 
thermodynamics of a stationary system with fixed $y$ are logically 
connected via the large deviation theory. 
This derivation shares the same spirit as
Helmholtz and Boltzmann's 1884 mchanical theory of heat \cite{gallavotti,campisi}: 
Both extend the notion of energy from a system with 
a fixed ``parameter'' $y$ to an entire family of systems 
with different $y$'s \cite{ma-qian-prsa}.

\section{An instantaneous deterministic energy balance equation}

	While the 
$\epsilon F\big[p_{\epsilon}(\vx,t)\big]\to\varphi\big(\hat{\vx}(t)\big)$
as $\epsilon\to 0$ and $p_{\epsilon}(\vx,t)\to\delta\big(\vx-\hat{\vx}(t)\big)$, the free energy dissipation, house-keeping heat, and 
entropy production rates \cite{ge-qian-10,qian-jmp}, also known as 
non-adiabatic, adiabatic, and total entropy production rates
\cite{esposito-vandenbroek}, become
\begin{subequations}

\begin{eqnarray}
	\epsilon f_{\mathrm{d}}\big[p_{\epsilon}(\vx,t)\big]  &\to& 
           \Big[\nabla \varphi(\vx) \cdot \mD(\vx)  \nabla \varphi(\vx)
            \Big]_{\vx=\hat{\vx}(t)},
\\
	\epsilon Q_{\mathrm{hk}}\big[p_{\epsilon}(\vx,t)\big] &\to&  
  \Big[ \vg(\vx) \cdot \mD^{-1}(\vx)\vg(\vx)
     \Big]_{\vx=\hat{\vx}(t)},
\\
   \epsilon e_{\mathrm{p}}\big[p_{\epsilon}(\vx,t)\big]  &\to&
  \Big[ \vb(\vx) \cdot \mD^{-1}(\vx)\vb(\vx)
     \Big]_{\vx=\hat{\vx}(t)}.
\end{eqnarray}
\label{3EP}
\end{subequations}
All three quantities are non-negative.  They are
linked through a Pythagorean-like equation, 
$\forall\vx\in\mathbb{R}^n$:
\begin{equation}
  \big\|\mD(\vx)\nabla \varphi(\vx) \big\|^2 +
   \big\| \vg(\vx)\big\|^2 =
      \big\| \vb(\vx) \big\|^2,
\label{p-like-eq}
\end{equation}
under the inner product $\langle \vu,\vv\rangle \equiv
\vv\mD^{-1}\vu$ and thus $\|\vu\|^2 = \vu\mD^{-1}\vu$.\footnote{This Pythagorean relation actually exists for 
stochastic $f_{\mathrm{d}},Q_{\mathrm{hk}},e_{\mathrm{p}}$ 
with finite $\epsilon$ if one defines inner product 
$$\langle \vu(\vx), \vv(\vx)\rangle\equiv \epsilon^{-1}\int_{\mathbb{R}^n}
\vu(\vx)\mD^{-1}(\vx)\vv(\vx)p_{\epsilon}(\vx,t)\rd\vx,
$$
and identifies $f_\mathrm{d}=\|\epsilon\mD(\vx)
\nabla\ln ( p_{\epsilon}(\vx,t)/\pi_{\epsilon}(\vx))\|^2$,
$Q_{\mathrm{hk}}=\|\vb(\vx)-\epsilon\mD(\vx)\nabla\ln\pi_{\epsilon}(\vx)\|^2$, and $e_\mathrm{p}=\| \vb(\vx)-\epsilon\mD(\vx)\nabla\ln p_{\epsilon}(\vx,t)\|^2$.
}

In the zero-noise limit, the deterministic motion follows
$\rd\vx(t)/\rd t = \vb(\vx)$; the balanace equation 
$\rd F/\rd t\equiv-f_{\mathrm{d}}=Q_{\mathrm{hk}}-e_{\mathrm{p}}$ now becomes
\begin{eqnarray}
    && \frac{\rd}{\rd t}\varphi\big(\vx(t)\big) 
      \ = \   \vb(\vx)\cdot \nabla\varphi(\vx)
\nonumber\\
   &=& \underbrace{ \vg(\vx) \cdot \mD^{-1}(\vx) \vg(\vx) }_{
         \text{ non-conservative pump } } - \underbrace { \vb(\vx)\cdot \mD^{-1}(\vx) \vb(\vx) }_{ \text{ energy dissipation} }.
\label{eq29}
\end{eqnarray}
Both terms before and after the minus sign in Eq. \eqref{eq29}
are non-negative. Eq. \eqref{eq29} constitutes a deterministic,
instantaneous energy balance law with the non-conservative 
pump as the source and energy dissipation as the sink,
 respectively.  This result provides a rigorous notion 
of ``energy'' for complex dynamics with a stochastic kinematic
description.  For a classical mechanical
system with potential force and friction, 
$\varphi$ is the sum of kinetic energy and potential
energy, and the energy dissipation is due to 
the friction.

\section{Perturbation theory of random processes and the higher-order approximation of $e_{\mathrm{p}}$}

From Eq. \eqref{3EP}, we have shown that the entropy production rate $e_{\mathrm{p}}$ is ``extensive'', i.e., $O\big(\epsilon^{-1}\big)$, and derived it's leading order as $\epsilon \rightarrow 0$. This leading order constitutes the part of entropy produced from the deterministic path of the ``dissipative dynamics" $\hat{\vx}'(t)=\vb(\hat{\vx})$. We can further obtain higher-order terms of the entropy production rate, which yields the part of entropy production around the deterministic trajectory due to the infinitesimal fluctuation.  We can visualize this by imaging that we follow the deterministic trajectory and measure the entropy production with a zoomed-in scale.  J. Keizer first realized this perspective 
provides a flucutation-dissipation theorem beyond
equilibrium \cite{keizer-acr,keizer}. 

From a mathematical standpoint, the dissipation part of entropy production is on the scale of the law of large number, while the fluctuation part is on the scale for the central limit theorem. This suggests us that the latter can be obtained by the perturbation of random processes with an appropriate scale.  Introducing
$\mathbf{Z}_\epsilon(t) =  \frac{1}{\sqrt{\epsilon}}(\vX_\epsilon(t)-\hat{\vx}(t))$ and following the usual 
perturbation approach \cite{freidlin-wentzell-book} with expansion
$\mathbf{Z}_\epsilon  =  \mathbf{Z}^{(0)}+ \sqrt{\epsilon}\mathbf{Z}^{(1)} + \cdots +\sqrt{\epsilon}^k \mathbf{Z}^{(k)} + \cdots$, we can write down a stochastic 
differential equation (SDE) for the $\mathbf{Z}^{(0)}(t)$
\begin{align}
    \rd \mathbf{Z}^{(0)}(t) = \mathbf{A}(\hat{\vx}(t)) \mathbf{Z}^{(0)}(t) \rd t + [2\mathbf{D}(\hat{\vx}(t))]^{\frac{1}{2}} \rd \vB(t), \label{time.inhom.sde}
\end{align}
where $\mathbf{A}(\vx)$ is the Jacobian matrix of $\vb(\vx)$. Eq. \eqref{time.inhom.sde} is a time-inhomogeneous linear SDE 
which can be solved as $\mathbf{Z}^{(0)}(t) \sim \mathcal{N}(0, \mathbf{\Sigma})$, where $\mathcal{N}$ represents Gaussian distribution and the covariance matrix $\mathbf{\Sigma}$ satisfies 
the equation
\begin{align} \label{dynamics.variance}
    \frac{\rd \mathbf{\Sigma}(t)}{\rd t} = \mathbf{A}(\hat{\vx}(t))\mathbf{\Sigma} + \mathbf{\Sigma}\mathbf{A}(\hat{\vx}(t))^T + 2 \mathbf{D}(\hat{\vx}(t)).
\end{align}
This equation under more restricted consideration had been 
obtained in \cite{gardiner,keizer}. In addition to the small noise expansion for the SDE, we can also expand the FPE of the scaled process $\mathbf{Z}_\epsilon$:
\begin{align} \label{fokkerplanck.exp}
    \hat{p}_\epsilon(\mathbf{z}, t) = \sum_{n=0}^\infty \hat{p}_n(\mathbf{z}, t)\sqrt{\epsilon}^n,
\end{align}
in which the $\hat{p}_0(\mathbf{z},t)$ corresponds exactly to the probability distribution of $\mathbf{Z}^{(0)}(t) \sim \mathcal{N}(0, \Sigma)$ \cite{gardiner}. By the change of variable $p_\epsilon(\vx,t) = \frac{1}{\sqrt{\epsilon}} \hat{p}_\epsilon(\mathbf{z}, t) $ and plugging it into the equation of entropy production rate in Eq. \eqref{3q}, we can obtain the higher-order approximation
\begin{eqnarray}
    \epsilon e_{\mathrm{p}}[p_\epsilon] &=& \big[  \vb(\vx)\cdot \mD^{-1}(\vx)\vb(\vx)\big]_{\vx=\hat{\vx}(t)} 
\\ 
    &+& \epsilon\big[\text{tr}\big(\mathbf{M}(\vx)\big) + 2  \vb(\vx) \cdot \mD(\vx)^{-1} \mathbf{m}\big]_{\vx=\hat{\vx}(t)} + o(\epsilon), 
\nonumber
\end{eqnarray}
where 
\begin{eqnarray}
    \mathbf{M}(\vx) &=& \mD(\vx)\mathbf{\Sigma}^{-1} + 2\mathbf{A}(\vx) + \mathbf{A}(\vx)^T\mD(\vx)^{-1} \mathbf{A}(\vx)\mathbf{\Sigma}  \nonumber \\
    &+& \vb(\vx) \cdot \mD(\vx)^{-1} \mathbf{H}(\vx) \mathbf{\Sigma} 
\end{eqnarray}
Note that $\mathbf{H}(\vx)$ is a rank $3$ tensor, for the vector $\mathbf{v} = \vb(\vx) \cdot \mD(\vx)^{-1}$,  $\mathbf{v}\mathbf{H}(\vx) = \sum_i v_i  \mathbf{H}_i(\vx)$, in which $\mathbf{H}_i(\vx)$ is the Hessian matrix of $b_i(\vx)$;  $\mathbf{\Sigma}$ follows Eq. \eqref{dynamics.variance}, and $\mathbf{m} = \int \mathbf{z}\hat{p}_1(\mathbf{z},t)\rd \mathbf{z}$.
Therefore, the $\epsilon$ order of the entropy production has the form
\begin{align}
    \underbrace{\text{tr}\big(\mathbf{M}(\hat{\vx}(t))\big)}_{\text{Gaussian fluctuation}} + \underbrace{2  \vb(\hat{\vx}(t))\cdot
    \mD(\hat{\vx}(t))^{-1} \mathbf{m}}_{\text{non-Gaussian fluctuation}},
\end{align}
where the first part involves $\mathbf{\Sigma}$ (the second moment with respect to the Gaussian distribution $\hat{p}_0(\mathbf{z},t)$) and the second part involves $\mathbf{m}$ (the first moment with respect to the next order $\hat{p}_1(\mathbf{z},t)$ in the expansion  \eqref{fokkerplanck.exp}). Interestingly, if $\mathbf{Z}_\epsilon(t)$ is a time-inhomogeneous Ornstein-Uhlenbeck process, then the non-Gaussian fluctuation part is always zero.  On the other hand, since the non-Gaussian  part is due to the nonlinearity of the vector filed $\vb$, this part of entropy production rate in the higher order exists uniquely in nonlinear dynamics, e.g. in limit cycles.

\section{Discussion}

In the theory of ordinary differential equations, Hamiltonian
dynamics with the conserved $H$ functions and gradient
systems with potential functions are two special classes that 
have been extensively studied \cite{perko}.  For a general
nonlinear dynamics $\dot{\vx}=\vb(\vx)$, it is not known
whether it always has an associated ``energetics''. 
Ventsel and Freidlin's large deviation theory revealed that
if $\vb(\vx)$ is the zero-noise limit of a noisy dynamics, 
a global quasi-potential function $\varphi(\vx)$ exists 
\cite{freidlin-wentzell,dykman,graham-tel}.   In fact,
 $\vb(\vx)=\mD(\vx)\nabla\varphi(\vx)+\vg(\vx)$,
where $\vg(\vx)\perp\nabla\varphi(\vx)$ at every $\vx$.
Graham and T\'{e}l further showed \cite{graham-tel-JSP}
$\nabla\cdot\big(\omega(\vx)\vg(\vx)\big)
=-\nabla\omega(\vx)\cdot\mD(\vx)\nabla\varphi(\vx)$,
where $\omega(\vx)$ represents a proper local measure at
$\vx$.  The present work shows that this system
of dynamic equations has a thermodynamic 
interpretation via a geometric equation:
$\|\vb\|^2 = \|\mD\nabla\varphi\|^2+\|\vg\|^2$,
corresponding to the instantaneous rates of total entropy 
production, free energy dissipation, and house-keep heat,
respectively.

	If one identifies $\dot{\vx}=\vb(\vx)$ as a kinematic description
of a complex dynamics, then the system in Eq. \eqref{hq-eq}
and Eq. \eqref{eq29} provide an energetic description that 
is hidden under the kinematics.  A few words concerning the
role of $\mD(\vx)$ are in order.
The concept of a gradient field on $\mathbb{R}^n$ requires
a notion of {\em distance}.  This is naturally provided by
the noise structure embedded in $\mD(\vx)$.  This is precisely
A. N. Kolmogorov's insights on the nature of probability
theory: One needs to have a probability given before 
carrying out probabilistic computations.

	For complex systems, not all ``stochasticity'' are due to
thermal noises.  In fact, the Mori-Zwanzig theory of projection 
operator clearly shows that \cite{mori-zwanzig} the dynamics
of a projection necessarily induces, in general, a stochastic term and a 
non-Markovian memory term.

	When $\vg=0$, the FPE in Eq. \eqref{eq1}
is a gradient flow in a proper mathematical space 
\cite{peletier}.  In terms of the Pythagorean-like equation in Eq.
\eqref{p-like-eq}, the stochastic dynamics following the Eq. \eqref{eq1} 
with $\vg=0$ has a maximal $f_{\mathrm{d}}$:  One leg of a ``triangle'' 
has the same length as the triangle's hypotenuse.
Thus, under an appropriate geometry, Onsager's {\em principle 
of maximum dissipation} can be generalized to nonlinear 
regime \cite{peletier-2}.  In recent years, 
the theory of nonequilibrium landscape has gained wider
recognitions in biological physics \cite{ao,qian-ge-mcb,hlzq}.
Since for a nonequilibrium stochastic system, its stationary 
state still has highly complex motions as NESS 
flux \cite{zqq,gqq,wang}, the $\varphi$ is only
half the story: the $\vg$ and its related $\omega$ 
provide the characterization of the NESS motion on each
and every $\varphi$ level set. 


\begin{acknowledgments}
We thank Jin Feng, Hao Ge, Yi-An Ma, Mark Peletier, and Jin Wang
for helpful discussions, and an anonymous reviewer for feedback.  
This work was supported in part by NIH R01 grants GM109964 and GM135396 (both PI: Sui Huang), and the Olga Jung Wan Endowed Professorship for the first author.
\end{acknowledgments}

\end{document}